\documentclass{JHEP}
\usepackage{epsfig}

\title{Computing the perturbative gluon condensate}
\author{G.\ Grunberg\thanks{Research
supported in part by the EC program ``Training and 
Mobility of 
Researchers'', Network ``QCD and Particle Structure'', contract 
ERBFMRXCT980194.}\\
        Centre de Physique Th\'eorique de l' Ecole  
Polytechnique (CNRS UMR C7644),\\
        91128 Palaiseau Cedex, France\\
        E-mail: \email{georges.grunberg@cpht.polytechnique.fr}}
\abstract{The normalization of the gluon condensate and of renormalon-related power corrections in QCD is
 computed under the assumption that their ``perturbative'' part dominates over any eventual extra contribution
from the non-trivial vacuum. The calculation is performed in the infrared finite coupling framework, assuming
an infrared fixed point  is present in the perturbative coupling down to low values
of $N_f$. The freezing perturbative coupling is reconstructed using a Banks-Zaks expansion approach.
Parameter-free predictions of  the low energy moments of the  coupling, which determine
the process-independent part of the power corrections, are obtained for a number of choices of the running
coupling.} 

\preprint{CPTh/S 025.0501}
\begin{document}

\section{Introduction}

Recent years have witnessed a rapid development of the area covering the borderline between ``perturbative'' and
``non-perturbative'' physics in QCD. In particular, perturbative ideas have been pushed increasingly far
towards the low-energy frontier  to deal with the phenomenology of power corrections. Still in these advances the normalization of 
power corrections is usually considered as an incalculable ``non-perturbative'' parameter, to be fitted from the data.
 This has been
the situation ever since their original  introduction \cite{SVZ} by Shifman, Vainshtein and Zakharov (SVZ).
 In this note, I suggest a possibility to compute
these parameters from first principles,  for the limited class of ``renormalon-related'' power corrections
 which include in particular the ``gluon condensate''. Among the various methods \cite{Ben} devised to deal 
with these contributions, the ``infrared finite coupling'' approach \cite{Dok-Web,DMW,Dok} stands out as a particularly 
attractive scheme. A framework where this approach can be justified has recently been suggested  \cite{Gru-conformal}, and the possibility
of a calculation of power corrections from {\em perturbative} input has  been  pointed out. The aim of this paper
is to implement the latter suggestion. After a brief review (section 2) of the infrared (IR) finite coupling approach and of the
proposal in \cite{Gru-conformal}, a method to construct the IR finite coupling from the Banks-Zaks
expansion is described in section 3. The results for the power corrections are given in section 4, and further discussed
in section 5 which also contains the conclusions.

\section{A framework for the IR finite coupling approach to power corrections}
In this approach the power corrections are parametrized in term of low energy moments of
 a ``non-perturbative''
coupling $\bar{a}=a+\delta a$ assumed to be IR finite, where $a$
($a\equiv {\alpha_s\over\pi}$) is the ``perturbative part'', and $\delta a$ a
``non-perturbative'' modification needed to make $\bar{a}$ IR finite. Although the approach
 can also deal
with Minkowskian quantities, consider as a simple example the case of an Euclidean observable $D(Q^2)$ in the
``single dressed gluon exchange approximation'' \cite{Gru-power,Gar-Gru-thrust}
\begin{equation}D(Q^2)=\int_0^{\infty}{dk^2\over k^2}\ \bar{a}(k^2)\
\Phi\left({k^2\over Q^2}\right)\label{eq:sdg-integral}\end{equation}
where $\Phi(k^2/Q^2)$, the ``loop momentum distribution function'' 
\cite{Ben},   is known \cite{Neu} from the relevant single dressed gluon diagrams. Introducing an
 IR cut-off $\mu_I$ to separate long and short distances, the right hand side is approximated
at large $Q^2$ by

\begin{equation}D(Q^2)\simeq c_n{\lambda_n(\mu_I)\over Q^{2n}}+\int_{\mu_I^2}^{\infty}{dk^2\over k^2}\ a(k^2)\
\Phi\left({k^2\over Q^2}\right)\label{eq:sdg-qlarge}\end{equation}
with $\lambda_n(\mu_I)=\int_0^{\mu_I^2}{dk^2\over k^2}\ \bar{a}(k^2)\ k^{2n}$,
where I assumed that  $\Phi(k^2/Q^2)\simeq c_n (k^2/Q^2)^n$ at low $k^2$ and the contribution of the $\delta a$
piece has been neglected above $\mu_I$. These steps can now be justified with the following two crucial assumptions:

i) The {\em perturbative} part $a$ of the coupling reaches a non-trivial IR fixed point at
low scales and is IR finite by itself, without the need for an
hypothetical $\delta a$ contribution. This statement is likely to be correct for $N_f$ slightly 
below 16.5 where the perturbative coupling has a Banks-Zaks fixed point \cite{BZ,White,G-bz}
beyond one-loop, and I assume it is still true down to $N_f=0$. This assumption is supported 
\cite{Stev,Gar-Kar,Gar-Gru-conformal} by 
the behavior
of the Banks-Zaks expansion for some QCD effective charges.

Actually, the previous statement must be correct within a range $N_f^*<N_f<16.5$
 which defines
the ``conformal window'' where the perturbative coupling is IR finite and causal \cite{Gru-short,Gru-conformal}.
Within the conformal window, there is by definition no $\delta a$ term, and we have

\begin{equation}D(Q^2)=D_{\overline{PT}}(Q^2)\label{eq:D-Dpt}\end{equation}
with

\begin{equation}D_{\overline{PT}}(Q^2)\equiv \int_0^{\infty}{dk^2\over k^2}\ a(k^2)\
\Phi\left({k^2\over Q^2}\right)\label{eq:sdg-pt-integral}\end{equation}
At large $Q^2$, one obtains as in eq.(\ref{eq:sdg-qlarge})

\begin{equation}D_{\overline{PT}}(Q^2)\simeq {C_{PT}(\mu_I)\over Q^{2n}}+\int_{\mu_I^2}^{\infty}{dk^2\over k^2}\ a(k^2)\
\Phi\left({k^2\over Q^2}\right)\label{eq:sdg-pt-qlarge}\end{equation}
with the normalization of the power correction

\begin{equation}C_{PT}(\mu_I)=c_n \int_0^{\mu_I^2}{dk^2\over k^2}\ a(k^2)\ k^{2n}\label{eq:C-pt}\end{equation}
given by a low energy moment of the {\em perturbative} coupling.
Since the latter is no more causal below $N_f^*$ (even if
it still IR finite there), eq.(\ref{eq:D-Dpt})
cannot be correct anymore for $N_f<N_f^*$, where the ``conformal window amplitude'' $D_{\overline{PT}}(Q^2)$
is expected to have unphysical Landau singularities in the (complex) $Q^2$ plane. We must therefore
have

\begin{equation}D(Q^2)=D_{\overline{PT}}(Q^2)+D_{\overline{NP}}(Q^2)
\label{eq:D-pt-np}\end{equation}
where\footnote{If $D_{\overline{PT}}(Q^2)$ is interpreted as the analytic
continuation in $N_f$ of the full conformal window amplitude, the decomposition eq.(\ref{eq:D-pt-np}) is general
\cite{Gru-short,Gru-conformal} and valid beyond the single dressed gluon approximation of
eq.(\ref{eq:sdg-integral}).} the ``genuine non-perturbative piece'' $D_{\overline{NP}}(Q^2)$ cancels the Landau
singularities present in $D_{\overline{PT}}(Q^2)$.  In the standard IR finite coupling approach  this piece would
correspond to the contribution of the
$\delta a$ part of the coupling in eq.(\ref{eq:sdg-integral}). Since the existence of
such a term is quite hypothetical, I shall {\em not} assume that the $D_{\overline{NP}}(Q^2)$ piece is  related to
 a (universal) non-perturbative QCD coupling. Still at large $Q^2$ this piece may contribute a
 ``non-perturbative component'' $C_{NP}$ to the ${\cal O}(1/Q^{2n})$ power correction

\begin{equation}D_{\overline{NP}}(Q^2)\simeq {C_{NP}\over Q^{2n}}\label{eq:C-np}\end{equation} 
so that below $N_f^*$ we have 

\begin{equation}D(Q^2)\simeq D_{\overline{PT}}(Q^2)+{C_{NP}\over Q^{2n}}\label{eq:NP-cond}\end{equation}
hence

\begin{equation}D(Q^2)\simeq {C(\mu_I)\over Q^{2n}}+\int_{\mu_I^2}^{\infty}{dk^2\over k^2}\ a(k^2)\
\Phi\left({k^2\over Q^2}\right)\label{eq:sdg-pt-qlarge}\end{equation}
with

\begin{equation}C(\mu_I)=C_{PT}(\mu_I)+C_{NP}\label{eq:C-pt-np}\end{equation}

ii) The second crucial assumption I shall make is that the ``non-perturbative'' component $C_{NP}$ can
in fact be neglected (for not too small $\mu_I$) in eq.(\ref{eq:C-pt-np}). This assumption, which actually
takes the exact counterpart of the   SVZ hypothesis \cite{SVZ} that the ``genuine non-perturbative piece''
$C_{NP}$ dominates over the ``perturbative'' fluctuations,
 can be
justified
\cite{Gru-conformal} in a number of ways. One is to observe that $C_{NP}$, which vanishes identically
for $N_f^*<N_f<16.5$ within the conformal window, may still be small for $N_f<N_f^*$ below
the conformal window, provided $N_f$ is close enough to $N_f^*$.
In \cite{Gru-short,Gru-conformal}, it was found that in fact $4<N_f^*<6$, which makes it at least plausible the neglect
of $C_{NP}$ at the ``real life'' QCD value $N_f=3$. Another (more drastic)
 possibility is that the power corrections
in $D_{\overline{PT}}(Q^2)$ and $D_{\overline{NP}}(Q^2)$ do not match (even though at low $Q^2$
the two components cancel their mutual Landau singularities below the conformal window), i.e. that the
 power corrections
are either entirely ``perturbative''  and contribute only to
$D_{\overline{PT}}(Q^2)$, or entirely ``non-perturbative''  and
contribute only to
$D_{\overline{NP}}(Q^2)$. This would mean that    $C_{NP}\equiv 0$ even {\em below} $N_f^*$, and only the
$C_{PT}$ component is present, for those condensates
(like the gluon condensate) which do not vanish
within the conformal window, whereas $C_{NP}\neq 0$ below $N_f^*$ only for those
condensates (like the quark condensate) which vanish identically within the conformal window,
 and therefore have no $C_{PT}$ component. In such a case, the neglect of $C_{NP}$ would be justified
at all $N_f$'s for the ``conformal window type'' of power corrections. Anyway, the working
hypothesis in the following shall be that one can compute the bulk of the latter type of power corrections
from eq.(\ref{eq:C-pt}) alone. In this way, the IR finite coupling approach  not only finds a natural framework, but
its predictiveness is enhanced since there is not any more any ``non-perturbative'' free parameter 
and the normalization of power corrections can be computed, as we demonstrate in the next section
(in this sense the approach goes beyond the operator product expansion even  when applied to Euclidean
quantities).

\section{Reconstructing the IR finite perturbative coupling: a Banks-Zaks expansion approach}

Even though the perturbative coupling appears to have an IR fixed point for large enough
$N_f$ beyond two-loop, this is not always manifest when one decreases
$N_f$. For instance, the Banks-Zaks fixed point at two-loop relies on having $\beta_1<0$, which
is not realized for $N_f<8$. Then one might rescue the fixed point with a negative three-loop
term, but even this feature is usually lost  at $N_f=3$. On the other hand, as mentioned in 
section 2, the Banks-Zaks expansion does signal in a number of cases the persistence of the fixed
point even down to $N_f=2$. This observation suggests the following strategy: try to reconstruct
the IR finite coupling, and eventually supply the missing higher order terms in
 the beta function $\beta(a)$,
 given the IR fixed point Banks-Zaks expansion. This  is an expansion  in powers of
 the distance 
$16.5-N_f$
from the top of the conformal window, which is proportional to $\beta_0$. The solution
$a^*=a^*(\epsilon)$ of the equation

\begin{equation}\beta(a)=\beta(a,\epsilon)=-\beta_0\ a^2-\beta_1\ a^3-\beta_2\ a^4-\beta_3\ a^5+...=0
\label{eq:fixed-point}\end{equation}
in the limit  $\beta_0\rightarrow 0$, with $\beta_i$ ($i\geq 1$) finite is obtained as a power series

\begin{equation}a^*=a^*(\epsilon)=\epsilon+\delta_1\ \epsilon^2+\delta_2\ \epsilon^3+...
\label{eq:BZ}\end{equation}
where

\begin{eqnarray}\delta_1&=&\beta_{1,1}-{\beta_{2,0}\over\beta_{1,0}}\nonumber\\
\delta_2&=&\delta_1^2+g_1\ \delta_1+\beta_{2,1}-g_2
\label{eq:coeff}\end{eqnarray}
The expansion parameter \cite{G-bz}  is  $\epsilon\equiv {8\over 321}(16.5-N_f)=
-{\beta_0\over \beta_{1,0}}$. The $\beta_{i,j}$, which are 
$N_f$-independent (but scheme dependent for $i>1$), are defined by
$\beta_1=\beta_{1,0}+\beta_{1,1}\ \beta_0$ ($\beta_{1,0}=-{107\over 16}$, $\beta_{1,1}={19\over 4}$),
$\beta_2=\beta_{2,0}+\beta_{2,1}\ \beta_0+\beta_{2,2}\ \beta_0^2$ (I assume $\beta_2$ is at most
 quadratic in $N_f$, hence in $\beta_0$) and  $g_1$, $g_2$ are given in eq.(\ref{eq:coeff-exp}). Given the knowledge of
 the 3-loop beta function
in (e.g.) the $\overline{MS}$ scheme, $\beta_{2,0}$ can be obtained \cite{G-BLM} from a one-loop
calculation of $a$ (see eq.(\ref{eq:b20})). I shall also use
 the related expansion  for the critical
 exponent 

\begin{eqnarray}\gamma&=&\gamma(\epsilon)={\partial\beta\over \partial a}(a^*,\epsilon)\label{eq:exponent}\\
&=&-\beta_{1,0}\ \epsilon^2(1+g_1\ \epsilon+
g_2\ \epsilon^2+...)\label{eq:exponent-BZ}
\end{eqnarray}
where 

\begin{eqnarray}g_1&=&\beta_{1,1}\nonumber\\ 
g_2&=&g_1^2+{\beta_{3,0}\over\beta_{1,0}}-\left({\beta_{2,0}\over\beta_{1,0}}\right)^2
\label{eq:coeff-exp}\end{eqnarray}
The $g_i$'s are scheme independent \cite{G-bz}, and \cite{Stev} $g_2=-8.89$.

The method relies on the differential equation \cite{G-bz} for $a^*(\epsilon)$

\begin{equation}{\partial\over\partial\epsilon}[\epsilon\ \nu(a^*,\epsilon)]=
\gamma(\epsilon)\ {da^*\over d\epsilon}\label{eq:equadiff-fixed-point}\end{equation}
where $\nu(a,\epsilon)$ is the $N_f$ dependent part of the beta function, after splitting
off the $\epsilon=0$ (i.e. $N_f=16.5$) piece $\beta(a,0)$

\begin{equation}\beta(a,\epsilon)\equiv \beta(a,0)-\epsilon\ 
\nu(a,\epsilon)\label{eq:beta-split}\end{equation}
Its  expansion in powers of $a$ is

\begin{equation}\nu(a,\epsilon)=-\beta_{1,0}\ a^2[1+\beta_{1,1}a+
(\beta_{2,1}+\beta_{2,2}\ \beta_0)a^2+...]\label{eq:nu}\end{equation} 
 Eq.(\ref{eq:equadiff-fixed-point}) follows by taking the total derivative with respect to $\epsilon$
of the relation $\beta(a^*,\epsilon)=0$ which defines the fixed point $a^*(\epsilon)$

\begin{equation}{\partial\beta\over\partial a}(a^*,\epsilon)\ {da^*\over d\epsilon}+
{\partial\beta\over\partial\epsilon}(a^*,\epsilon)=0\label{eq:deriv}\end{equation}
and using  eq.(\ref{eq:exponent})
and ${\partial\beta\over\partial\epsilon}(a,\epsilon)=-{\partial\over\partial\epsilon}[\epsilon\ \nu(a,\epsilon)]$
(eq.(\ref{eq:beta-split})).

It is convenient to introduce the function $\epsilon^*(a)$, which is the inverse of the 
Banks-Zaks function $a^*(\epsilon)$: for given $a$, $\epsilon^*(a)$ is the value
of $\epsilon$ (i.e. of $N_f$) where $\beta(a,\epsilon)=0$. The knowledge of $\epsilon^*(a)$
and of $\nu(a,\epsilon)$  determine $\beta(a,0)$, hence the full beta function. Indeed
using eq.(\ref{eq:beta-split})
the condition $\beta[a,\epsilon^*(a)]=0$ becomes

\begin{equation}\beta(a,0)= \epsilon^*(a)\ \nu[a,\epsilon^*(a)]
\label{eq:beta0}\end{equation}
Hence

\begin{equation}\beta(a,\epsilon)=\epsilon^*(a)\ \nu[a,\epsilon^*(a)]-\epsilon\ 
\nu(a,\epsilon)\label{eq:beta-fp}\end{equation}
In term of $\epsilon^*(a)$ eq.(\ref{eq:equadiff-fixed-point}) reads

\begin{equation}{\partial\over\partial\epsilon}[\epsilon\ \nu(a,\epsilon)]\vert_{\epsilon=\epsilon^*}
={\gamma(\epsilon^*)\over {d\epsilon^*\over da}}\label{eq:equadiff1-fixed-point}\end{equation}
 Eq.(\ref{eq:equadiff1-fixed-point}) gives a constraint on $\nu(a,\epsilon)$
given the Banks-Zaks functions $\gamma(\epsilon)$ and $a^*(\epsilon)$. This constraint
is not sufficient to determine $\nu(a,\epsilon)$ (and the beta function) without further assumptions.
In the following I shall assume that
$\nu(a,\epsilon)=\nu_0(a)$ is independent of $\epsilon$, i.e. that the beta function coefficients
are at most linear in $N_f$ (or $\beta_0$): this amounts to an approximation, in the spirit of the Banks-Zaks approach,
where one keeps only the leading $\epsilon=0$ term in an expansion of $\nu(a,\epsilon)$ 
in powers of $\epsilon$ (in particular, one neglects
the $\beta_{2,2}\ \beta_0$ term in eq.(\ref{eq:nu})). Then eq.(\ref{eq:equadiff1-fixed-point}) gives

\begin{equation} \nu_0(a)=
{\gamma[\epsilon^*(a)]\over {d\epsilon^*\over da}}\label{eq:equa-nu}\end{equation}
and from eq.(\ref{eq:beta-fp}) one gets

\begin{equation}\beta(a,\epsilon)=[\epsilon^*(a) -\epsilon]\ {\gamma[\epsilon^*(a)]
\over {d\epsilon^*\over da}}\label{eq:beta1-fp}
\end{equation}
Using the Banks-Zaks expansions of the fixed point $a^*(\epsilon)$ and of the critical
exponent $\gamma(\epsilon)$ truncated to a given order as input, 
eq.(\ref{eq:beta1-fp})
yields a corresponding ``improved'' approximation to the beta function, which  displays a built-in fixed
point at $a=a^*(\epsilon)$.

In this approach, the leading order (LO) approximation thus gives $\epsilon^*(a)=a$ and 
$\gamma[\epsilon^*(a)]=-\beta_{1,0}\ a^2$.
The next-to-leading order (NLO)  approximation   uses the NLO Banks-Zaks expansions of the
 fixed point  and of the critical
exponent: $\epsilon^*(a)$
is then obtained by  inverting  eq.(\ref{eq:BZ}) (with $\delta_2=0$), i.e. solving for $\epsilon^*$ in
$a=\epsilon^*+\delta_1\ \epsilon^{*2}$
 and reporting in eq.(\ref{eq:beta1-fp}), with
 $\gamma(\epsilon^*)=-\beta_{1,0}\ \epsilon^{*2}(1+g_1\ \epsilon^*)$. 
 The next-to-next-to-leading order (NNLO)  approximation uses the NNLO Banks-Zaks expansions of 
the fixed point  (which requires  the knowledge of $\beta_{2,1}$) and of the critical
exponent (eq.(\ref{eq:BZ}) and (\ref{eq:exponent-BZ})), etc...

The approximation can be further systematically 
 improved by including the knowledge of the known $N_f$-dependent
terms in the beta function. For instance, if the three-loop $\beta_2$ coefficient
is known, one can include the knowledge\footnote{Actually, given that $0<a<a^*= \cal {O}(\epsilon)$, this term 
is effectively of the same order as the $\beta_{3,1}\ a^5$ term in $\nu(a,\epsilon)$ (eq.(\ref{eq:nu})), and should
be taken as input only together with the latter, i.e. at the NNNLO level.} of the
term quadratic in $\beta_0$ in $\beta_2$  with the ansatz
\begin{equation}\nu(a,\epsilon)=\nu_0(a)+\beta_{1,0}^2\ \beta_{2,2}\ a^4\ \epsilon\label{eq:ansatz-nu}\end{equation}
 where $\nu_0(a)\equiv\nu(a,0)$
is independent of $\epsilon$ (the knowledge of  $\beta_{2,0}$ and $\beta_{2,1}$ is contained in the NLO and NNLO terms
 in the Banks-Zaks expansion of $a^*$, as mentioned above). 
 Eq.(\ref{eq:equadiff1-fixed-point}) then fixes $\nu_0(a)$ from

\begin{equation} \nu_0(a)+2\ \beta_{1,0}^2\ \beta_{2,2}\  a^4\ \epsilon^*(a)=
{\gamma[\epsilon^*(a)]\over {d\epsilon^*\over da}}\label{eq:equa-nu1}\end{equation}
which yields $\nu(a,\epsilon)$, hence from eq.(\ref{eq:beta-fp})

\begin{equation}\beta(a,\epsilon)=[\epsilon^*(a) -\epsilon]\ {\gamma[\epsilon^*(a)]
\over {d\epsilon^*\over da}}-\beta_{1,0}^2\ \beta_{2,2}\ a^4\ [\epsilon^*(a) -\epsilon]^2\label{eq:beta2-fp}
\end{equation}

\section{Results}
The ``coupling'' appearing in eq.(\ref{eq:sdg-pt-integral}) should be viewed as a physical, gauge-independent
quantity, just as the observable $D(Q^2)$ to which it is directly related. In the IR finite coupling
approach, it is also assumed to be universal, i.e. the same for all observables. The existence of such
an object is still speculative. It is attractive to identify this coupling to the ``skeleton coupling''
\cite{Gru-power,Gar-Gru-thrust,Brodsky}
associated to a (yet hypothetical) ``QCD skeleton expansion''. A promising approach in this direction
is provided by the ``pinch technique'' construction \cite{Watson,Papavassiliou}. The pinch coupling
is presently known only at one-loop, where it is related to the $\overline{MS}$ coupling by

\begin{equation}a(k^2)=a_{\overline{MS}}(\mu^2)+\left[-\beta_0 \left(\log(k^2/\mu^2)-5/3\right)+d_1\right]\ a_{\overline{MS}}^2(\mu^2)+... 
\label{eq:oneloop}\end{equation}
with $d_1\vert pinch=1$.
An alternative suggestion \cite{Dok-Web} is to use the ``gluon bremsstrahlung
coupling'' \cite{CMW}, also known to the one-loop level eq.(\ref{eq:oneloop}) with $d_1\vert brems=1-\pi^2/ 4$.
Since the full three-loop beta function coefficient (hence $\delta_2$) is not yet known for these two couplings,  I shall apply
the method of section 3 in the NLO 
approximation described there. Actually, since the Banks-Zaks expansion
of the critical exponent is  known \cite{Stev} up to NNLO
 (eq.(\ref{eq:exponent-BZ})), and may be reliable \cite{Gru-conformal}
even down to $N_f=3$,
 I shall go half-way towards the NNLO approximation, and   use eq.(\ref{eq:exponent-BZ})
 in eq.(\ref{eq:beta1-fp}), while still using eq.(\ref{eq:BZ}) (with $\delta_2=0$) to fix 
$\epsilon^*(a)$.
 The input scheme dependent numerical
 values following from eq.(\ref{eq:oneloop}) and the relation \cite{G-BLM}

\begin{equation}d_1=-{\beta_{2,0}-\beta_{2,0}^{\overline {MS}}\over\beta_{1,0}}\label{eq:b20}\end{equation}
are \cite{Brodsky}
${\beta_{2,0}\over\beta_{1,0}}\vert pinch=2.61$  and
 ${\beta_{2,0}\over\beta_{1,0}}\vert brems=2.61+{\pi^2\over 4}=5.08$. Hence
$\delta_1\vert pinch=2.14$ while $\delta_1\vert brems=-0.33$ (a smaller correction!). It follows 
from eq.(\ref{eq:BZ})
 that at $N_f=3$ the IR fixed point $a^*=0.299$ for the gluon bremsstrahlung coupling, smaller
 then the corresponding value $a^*=0.578$ for the pinch coupling which is subject to rather large
 uncertainties. 

As a third alternative, I would like to suggest the  ``universal coupling''
introduced in \cite{G-bz}, because of its simplicity. It is defined\footnote{I assume linear $N_f$
dependence. Otherwise there is the more general solution
$\beta(a,\epsilon)=\gamma(a)(a-\epsilon)+\gamma_1(a)(a-\epsilon)^2+...$.}
by the condition $\delta_i=0$ 
for all $i$'s, i.e. $a^*(\epsilon)\equiv \epsilon$, and therefore its beta function can be expressed
 entirely in term of the critical exponent

\begin{eqnarray}\beta(a,\epsilon)&=&(a-\epsilon)\gamma(a)\nonumber\\
&=&-(\beta_0\ a^2+\beta_{1,0}\ a^3)(1+g_1\ a+
g_2\ a^2+...)\label{eq:universal}
\end{eqnarray}
At $N_f=3$ the IR fixed point is $a^*=\epsilon=0.336$. The scale is fixed  knowing that
 ${\beta_{2,0}\over\beta_{1,0}}\vert universal=\beta_{1,1}$ (from $\delta_1=0$), which determines (eq.(\ref{eq:b20}))
 $d_1\vert universal=-1.137$, and the natural assumption that the term proportional to $\beta_0$ in eq.(\ref{eq:oneloop}) is the same.

The bremsstrahlung coupling beta function at $N_f=3$
is shown in Fig.1. Note the {\em negative} ultraviolet fixed point at $a\simeq -0.17$. It corresponds to a zero
of the critical exponent eq.(\ref{eq:exponent-BZ}) at $\epsilon\simeq -0.15$, and is a necessary condition for the scenario 
in \cite{Gru-conformal} to determine the bottom $N_f^*$ of the conformal window from the condition
$\gamma(\epsilon)=1$, which yields $N_f^*\simeq 4$ if one uses eq.(\ref{eq:exponent-BZ}).  
The resulting running coupling is shown in Fig.2, where I used $\alpha_s^{\overline{MS}}(M_Z)=0.117$ as
 input (eq.(\ref{eq:oneloop}) yields the corresponding input value of the bremsstrahlung coupling).

\EPSFIGURE{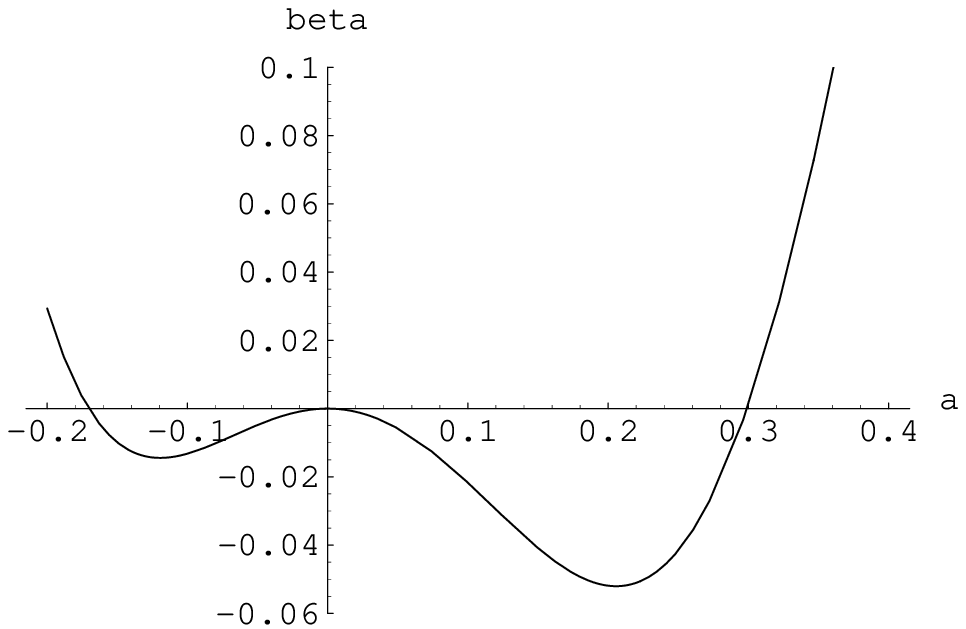}{The bremsstrahlung coupling beta function ($N_f=3$).}

\EPSFIGURE{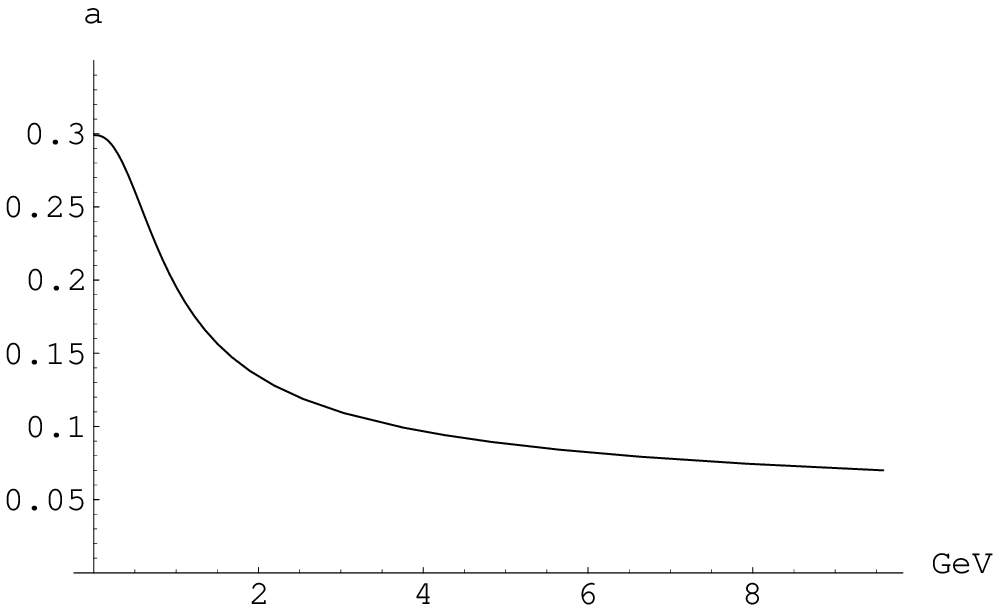}{The bremsstrahlung coupling  ($N_f=3$, $\alpha_s^{\overline {MS}}(M_Z)=0.117$).}

%\begin{figure}[H]
%\begin{center}
%\mbox{\kern-0.5cm
%\epsfig{file=gamma1.ps,width=10.0truecm,angle=0}}
%\end{center}
%\caption{The critical exponent as a function of $N_f$:
% top: ${\cal O}(\epsilon^3)$ order; middle: Pad\'e; bottom: ${\cal O}(\epsilon^4)$ order.}
%\label{pert}
%\end{figure}

It is then straightforward to compute the first few low energy moments of the coupling

\begin{equation}a_{2n-1}(\mu_I)\equiv \int_0^{\mu_I^2}n\ {dk^2\over k^2}\ \left({k^2\over
  \mu_I^2}\right)^n\ a(k^2)\label{eq:moments}\end{equation}
In term of the beta function they are given by

\begin{equation}a_{2n-1}(\mu_I)=\int_{a^*}^{a_I} n\ {da\over\beta(a)}\ a\ \exp[n\Phi(a,a_I)]
\label{eq:moments1}\end{equation}
where

\begin{equation}\Phi(a,a_I)\equiv \int_{a_I}^a  {dx\over\beta(x)}=\log\left({k^2\over\mu_I^2}\right)
\label{eq:RG}\end{equation}
is the solution of the renormalization group equation with $a_I\equiv a(\mu_I^2)$. 
Taking $\mu_I=2 GeV$, one gets the results in Table 1 if $\alpha_s^{\overline{MS}}(M_Z)=0.117$
and those in Table 2 if $\alpha_s^{\overline{MS}}(M_Z)=0.120$.  Note the sensitivity to the high
 energy input value of $\alpha_s$.

\TABULAR[r]{c|c|c|c}{  &bremsstrahlung&universal&pinch\\\hline\\$a_0$&0.207 (0.222)&0.225 (0.246) &0.330 (0.366) \\\hline
\\$a_1$&0.176 (0.198) &0.187 (0.212) &0.256 (0.300) \\\hline\\$a_3$&0.155 (0.173) &0.163 (0.183) &0.210 (0.243) }
{Moments for $\alpha_s^{\overline{MS}}(M_Z)=0.117$.}

  These results are subjected to  
theoretical uncertainties, stemming from the magnitude of the IR values of the coupling which should
 induce seizable
higher order corrections. The convergence of the Banks-Zaks expansion is  bad in the pinch coupling case (which
has a large IR value), and 
the knowledge of the 3-loop beta function coefficient and of $\delta_2$ is essential for a more reliable prediction. The 
situation looks better for the``universal coupling''and the bremsstrahlung coupling. To assess the convergence of the
 expansion,  the results for the moments in the NLO approximation where one uses only
the first two terms in the Banks-Zaks expansion of $\gamma$ (eq.(\ref{eq:exponent-BZ}))
are quoted within parenthesis in the tables. 

\TABULAR[r]{c|c|c|c}{  &bremsstrahlung&universal&pinch\\\hline\\$a_0$&0.217 (0.232)&0.237 (0.259) &0.353 (0.390) \\\hline
\\$a_1$&0.188 (0.213) &0.202 (0.230) &0.281 (0.332) \\\hline\\$a_3$&0.167 (0.189) &0.176 (0.201) &0.233 (0.273) }
{Moments for $\alpha_s^{\overline{MS}}(M_Z)=0.120$.}

 The $n=0$ moment gives the  
process-independent part of the
normalization of the $1/Q$ power corrections. If one uses the gluon bremsstrahlung ansatz for the coupling, 
the predicted value
 is in qualitative agreement  with the experimentally determined \cite{Dok} one  ($a_0\simeq 0.14-0.17$),
 although it should be remembered that the latter depends on the way the ``perturbative part'' of the
amplitude (the piece above $\mu_I$ in eq.(\ref{eq:sdg-pt-qlarge})) is handled, as well as upon extra
assumptions in the case of non-inclusive Minkowskian observables. The $n=3$ moment gives the normalization
of the ``gluon condensate''

\begin{equation}<{\alpha_s\over\pi}G^2>_{\mu_I}={3\over 2\pi^2}a_3(\mu_I)\mu_I^4\label{Gluon-cond}\end{equation}
Note the definition used here involves an arbitrary IR cut-off $\mu_I$, as
necessary in the case of renormalon-related power corrections. If one
wants to compare\footnote{I am indebted to Al. Mueller for raising the question.} to the effective 
 phenomenological SVZ
 definition  \cite{SVZ}, one can just
compute the integral in eq.(\ref{eq:sdg-pt-integral}) (which does not depend on $\mu_I$) for any given
Euclidean observable where the gluon condensate gives the leading power correction, and fit the result with the SVZ
 ansatz. For instance, for the Adler D function
 
\begin{equation}D(Q^2)\simeq a(Q^2)+{2\pi^2\over 3}{1\over Q^4}<{\alpha_s\over\pi}G^2>\label{G-SVZ}\end{equation}
where the ${2\pi^2\over 3}$ factor is the leading order coefficient function. Similarly, the SVZ condensate
$<{\alpha_s\over\pi}G^2>$ could be defined from the basic observable
$a_3(\mu_I)$ (eq.(\ref{eq:moments}) with n=2), where the IR cut-off $\mu_I$ now plays the role of the high energy
scale $Q$, by

\begin{equation}a_3(\mu_I)\simeq a(\mu_I^2)+{2\pi^2\over 3}{1\over
\mu_I^4}<{\alpha_s\over\pi}G^2>\label{G-eff}\end{equation}
For $\mu_I=2 GeV$,  eq.(\ref{G-eff}) yields $<{\alpha_s\over\pi}G^2>\simeq 0.05\ GeV^4$ for
$\alpha_s^{\overline{MS}}(M_Z)=0.117$,  which looks reasonable compared to the standard SVZ value. However, this 
comparison is actually devoid of significance due to the following intriguing fact: varying the scale $\mu_I$
in eq.(\ref{G-eff}), one finds the discrepancy $a_3(\mu_I)- a(\mu_I^2)$ between $a_3(\mu_I)$ and its lowest order
perturbative approximation $a(\mu_I^2)$ decreases {\em much slower} then the inverse fourth power of $\mu_I$! A
similar result is obtained if one uses eq.(\ref{G-SVZ}) (with the ``loop momentum distribution function''
$\Phi(k^2/Q^2)$
 taken from \cite{Neu}). Since the (principal value regulated) Borel sum of the perturbative series associated
to the observables $a_3(\mu_I)$ (or  $D(Q^2)$) are known \cite{Grunberg-FP, D-U} to differ from the exact values by
just such an
 ${\cal O}(1/\mu_I^4)$ (resp. ${\cal O}(1/Q^4)$) correction, one is bound to conclude that the naive treatment of
approximating the Borel sum by its leading order term does not work\footnote{Similar results are obtained 
if one uses \cite{Gar-Gru-thrust} the BLM scale \cite{Brodsky} in $a$.}
here. This is another point of discrepancy with the standard SVZ procedure, on top of the assumption that
the ``perturbative part'' of the condensate dominates.

\section{Discussion and conclusions}
The essential assumption in the present approach is that the {\em perturbative}  beta function has
an IR fixed point at least down to $N_f=3$. This is partly implemented by constructing beta functions
with {\em negative} three-loop coefficients: at NNLO the method of section 3 yields
 $\beta_2=\beta_{2,0}+\beta_{2,1}\ \beta_0$ where both $\beta_{2,0}$ and $\beta_{2,1}$ turn out negative
(see footnote 4)
for the considered couplings. Actually,  essentially 
the same results can be obtained (at least for the bremsstrahlung coupling)
 in a simpler
way, which makes it transparent the reason for the existence of the IR fixed point.
Indeed, consider the 4-loop beta function eq.(\ref{eq:fixed-point}), and observe that in the IR region 
the usual power counting should be modified: namely, given that $a$ is ${\cal O}(\beta_0)$ there, to ${\cal O}(a^6)$
accuracy one should drop the ${\cal O}(\beta_0^2)$ term in $\beta_2$, and keep only the
leading ${\cal O}((\beta_0)^0)$ term
 in $\beta_3$, i.e. use the effective  4-loop beta function (in accordance with the remark in footnote 1)

\begin{equation}\beta_{eff}(a,\epsilon)=-\beta_0\ a^2-\beta_1\ a^3-(\beta_{2,0}+\beta_{2,1}\ \beta_0) a^4-
\beta_{3,0}\ a^5+{\cal O}(a^6)
\label{beta-eff}\end{equation}
(in the ultraviolet region, this beta function has  of course only the ${\cal O}(a^4)$ accuracy
 of the 2-loop beta function). In the case of the bremsstrahlung coupling, the results obtained using the 4-loop
 $\beta_{eff}$ turn out to be very close to those of
 section 4
in the NNLO approximation. For instance $\beta_{eff}$ has an
 IR fixed point at $a^*=0.294$ if $N_f=3$ (I used $\beta_{3,0}=37.76$ from eq.(\ref{eq:coeff-exp})),
 and one gets: $a_0=0.201$, $a_1=0.177$ and $a_3=0.156$ if $\alpha_s^{\overline{MS}}(M_Z)=0.117$,
 and $a_0=0.210$,
 $a_1=0.189$ and $a_3=0.168$ if $\alpha_s^{\overline{MS}}(M_Z)=0.120$. Similarly in NLO one should use a 3-loop
$\beta_{eff}(a,\epsilon)=-\beta_0\ a^2-\beta_1\ a^3-\beta_{2,0}\ a^4+{\cal O}(a^5)$, and in LO a 2-loop
$\beta_{eff}(a,\epsilon)=-\beta_0\ a^2-\beta_{1,0}\ a^3+{\cal O}(a^4)$. The presence of an IR fixed point
in $\beta_{eff}$ down to low values of $N_f$ seems to be a general phenomenon, at least up to NLO. This is obvious in LO,
 since $\beta_{1,0}$
is scheme independent, but less so  in NLO where $\beta_{2,0}$ is scheme dependent. Nevertheless it
turns out that $\beta_{2,0}$ is {\em negative} for all known physical effective charges \cite{Brodsky},
 as well as for the three couplings quoted above. Consequently, there may be a positive zero in the 3-loop
$\beta_{eff}$ correctly signalling an IR fixed point, even if the standard 3-loop beta function  has no positive zero with all its coefficients of
 the same sign.

 At NNLO, the presence of an IR fixed point in the 4-loop $\beta_{eff}$ may be jeopardized
by large positive values of $\beta_{2,1}$ and (or) $\beta_{3,0}$. Actually, $\beta_{2,1}$ turns out to
be  negative for all known\footnote{For the pinch coupling and the bremsstrahlung coupling, $\beta_{2,1}$ has
been ``predicted'' from the assumption that $\delta_2\simeq 0$, which yields (eq.(\ref{eq:coeff}))
 $\beta_{2,1}=-23.6$ for the pinch coupling and
  $\beta_{2,1}=-7.43$ for the bremsstrahlung coupling. This assumption turns out to yield rather
good results in the case of the effective charges associated to the Adler D-function and the polarized
($g_1$) and non-polarized ($F_1$)  Bjorken sum rules, for which the ``predicted'' values are respectively
 $\beta_{2,1}=-16.17, -9.98, -6.97$   compared to the exact values (corrected for some 
numerical inaccuracies in \cite{Brodsky})
 $\beta_{2,1}=-15.94, -11.19, -6.81$. The partial reason for this success 
are the large cancellations between $g_2$ (the ``scheme independent'' contribution to $\delta_2$ 
in eq.(\ref{eq:coeff})) and the ``scheme dependent'' contribution which involves $\delta_1$ and $\beta_{2,1}$.} 
effective charges (except the one (``$a_V$'') defined by the static QCD potential,
 where it is 
 positive \cite{Brodsky} but small enough not to destabilize the  fixed point). The real problem comes from
the 4-loop coefficient
$\beta_{3,0}$, which is {\em positive} for all known effective charges (except again $a_V$, where it is
negative and tiny). In the case of the pinch coupling, it turns out in fact too large (one gets $\beta_{3,0}=164.7$
from eq.(\ref{eq:coeff-exp})) for the 4-loop $\beta_{eff}$ to have
an IR fixed point  if $N_f<13$. Similarly, in the case of the Adler  D-function effective charge where
$\beta_{3,0}=127$, the 4-loop $\beta_{eff}$ does not have an IR fixed point if $N_f<11$. For all other
 effective
 charges however the 4-loop $\beta_{eff}$ does exhibit an IR fixed point    down to $N_f=0$! However, in those 
cases of large positive $\beta_{3,0}$ (which is a consequence of a {\em small} $\beta_{2,0}$, 
see eq.(\ref{eq:coeff-exp})), the method of section 3 provides an effective resummation of the relevant higher
order terms, obtained under the assumption the Banks-Zaks expansions of the IR fixed
point and of the critical exponent do converge: {\em all} known effective charges then
 appear\footnote{This is even true for the effective
 charge associated \cite{Vermaseren} to Higgs decay.
 In this case
however one gets a large fixed point value $a^*={\cal O}(1)$, and convergence of the Banks-Zaks expansion is
doubtful for $N_f<4$.}
 after resummation to have an IR fixed point down 
to $N_f=0$ (although the convergence of the fixed point Banks-Zaks expansion becomes  problematic already
at $N_f=3$  for some of them, such as the pinch coupling).

The suggestion of perturbative freezing of the coupling at low $N_f$ was first made in \cite{Stev}. There is
however an essential difference with the present proposal: it is {\em not} suggested here that the perturbative
IR fixed point has anything to do with the low energy behavior of the full QCD amplitudes below the conformal
window, which is entirely non-perturbative. For instance, as observed in \cite{Rafael}
 spontaneous chiral symmetry breaking considerations
at large $N_c$ imply 
 the Adler D-function must vanish
  at zero momentum, which is inconsistent with the positive value expected from perturbative freezing. What is
 suggested instead is that the perturbative freezing at low $N_f$ is relevant to determine the normalization of 
renormalon-related condensates and power corrections which appear in the {\em short distance} expansion
of amplitudes. This amounts to the recognition that objects like the ``gluon condensate'', at the
difference of the quark condensate, are of a basically 
``perturbative'' nature, and thus unrelated to ``genuine'' non-perturbative properties of the vacuum 
such as  chiral symmetry breaking or confinement.
The notion of a ``conformal window'' is an essential 
part of the present proposal: only those power corrections which are already present within the conformal
window are amenable to a perturbative treatment, and below the conformal window there are other
really ``non-perturbative'' contributions which are crucial to determine the true low energy properties of QCD.
Moreover, it was shown in \cite{Gru-conformal} that the assumption that the perturbative IR fixed point
persists below the bottom
$N_f^*$ of the conformal window  leads to the condition $\gamma(\epsilon)=1$ to determine 
$N_f^*$. It is interesting that this condition gives
$N_f^*\simeq 4$, rather close to the ``real life'' QCD value $N_f=3$, which might give an alternative
justification to the suggested calculation procedure based on the ``anti-SVZ'' hypothesis that the ``perturbative''
piece of the condensates actually dominates over the ``non-perturbative'' fluctuations. Note that the opposite SVZ
assumption of dominance of the non-perturbative piece has been questioned previously in the literature (see e.g.
\cite{Ben}). Furthermore, even if the present assumption
 turns out to be invalid, the results of this paper are still useful to extract from experiment the
``truly non-perturbative'' part $C_{NP}$, which is given a completely unambiguous definition through
eq.(\ref{eq:NP-cond}).

The typical example of a ``perturbative'' conformal window amplitude is the ``single dressed gluon'' integral of
 eq.(\ref{eq:sdg-pt-integral}), where the  running coupling inside the integral is IR finite, and
calculable from {\em perturbative} input through an (eventually resummed) Banks-Zaks expansion. It is implicitly
assumed that this particular coupling is free of IR renormalons and can be unambiguously determined from its
perturbative series (say, by Borel summation). The corresponding Banks-Zaks series should then be also
 Borel summable.
Note also that the integral eq.(\ref{eq:sdg-pt-integral}) is free of any renormalon 
ambiguity, although renormalons are  present in the corresponding perturbative series, but is still
expected to be affected below the conformal window
by unphysical Landau singularities in the complex $Q^2$ plane. Such an amplitude \cite{Grunberg-FP,D-U} represents a natural form
of a generalized perturbation theory, which gives the background on top of which genuine
non-perturbative contributions may take place below $N_f^*$. The calculation of the ``perturbative condensate'',
although using only perturbative information, goes beyond a mere renormalon estimate,
 since there is usually a part \cite {Gru-condensate,Gru-conformal} of
the low momentum contribution of the perturbative coupling which is {\em not} 
 determined {\em only} \cite{Gru-condensate,G-qed}  by renormalons.
 The assumption that renormalon-related power corrections
are ``perturbative'' in the above sense  also
 gives a straightforward justification to the IR finite coupling
approach to power corrections, and leaves no arbitrary free parameter (except of course the overall QCD scale)
to be fixed from experiment. The main conceptual problem in this framework remains to find the identity of the
(hopefully unique)
perturbative IR finite QCD coupling which determines the power corrections, and derive the systematic form
 to all orders of the (yet to be constructed) generalized perturbation theory, perhaps \cite{Brodsky} along the
lines of a  QCD ``skeleton expansion'': this is however a problem of a basically perturbative nature.

\acknowledgments
I thank Yu.L. Dokshitzer and A.H. Mueller for stimulating  discussions. I also wish to thank the referee for constructive criticism.

%\newpage


\begin{thebibliography}{9}


\bibitem{SVZ} M.A. Shifman, A.I. Vainshtein and V.I. Zakharov, {\em Nucl. Phys.} {\bf B 147}  (1979) 385.


\bibitem{Ben} See e.g. M. Beneke, {\em Phys. Rept.} {\bf B 317} (1999) 1 [hep-ph/9807443].

\bibitem{Dok-Web} Yu.L. Dokshitzer and B.R. Webber, {\em Phys. Lett.} {\bf B 352} (1995) 451 [hep-ph/9504219].

\bibitem{DMW}  Yu.L. Dokshitzer, G. Marchesini and B.R. Webber,  
{\em Nucl. Phys.} {\bf B 469}  (1996) 93 [hep-ph/9512336].

\bibitem{Dok} For a review and further references, see  Yu.L. Dokshitzer, in 
{\em International Conference ``Frontiers of Matter''}, Blois, France, June 1999 
[hep-ph/9911299].

\bibitem{Gru-conformal} G. Grunberg, Conformal window and Landau singularities, hep-ph/0104098.

\bibitem{Gru-power} G. Grunberg, {\em JHEP} {\bf 11} (1998)
 006 [hep-ph/9807494].

\bibitem{Gar-Gru-thrust} E. Gardi and G. Grunberg, {\em JHEP} {\bf 11} (1999)
 016 [hep-ph/9908458]. 

\bibitem{Neu} M. Neubert, {\em Phys. Rev.} {\bf D 51} (1995) 5924 [hep-ph/9412265].


\bibitem{BZ} T. Banks and A. Zaks,  
{\em Nucl. Phys.} {\bf B 196} (1982) 189.

\bibitem{White} A.R. White, {\em Phys. Rev.} {\bf D 29} (1984) 1435  ; {\em Int. J. Mod. Phys.} {\bf A 8} (1993) 4755  [hep-th/9303053].

\bibitem{G-bz} G. Grunberg, {\em Phys. Rev.} {\bf D 46} (1992) 2228  .


\bibitem{Stev} P.M. Stevenson, {\em Phys. Lett.} {\bf B 331} (1994) 187 [hep-ph/9402276]; S.A. Caveny and P.M. Stevenson, 
hep-ph/9705319.

\bibitem{Gar-Kar} E. Gardi and M. Karliner,  
{\em Nucl. Phys.} {\bf B 529} (1998) 383
  [hep-ph/9802218].

\bibitem{Gar-Gru-conformal} E. Gardi and G. Grunberg, {\em JHEP} {\bf 03}   (1999) 024
 [hep-th/9810192]. 

\bibitem{Gru-short} G. Grunberg, Fixing the conformal window in QCD, hep-ph/0009272.

\bibitem{G-BLM} G. Grunberg, {\em Phys. Lett.} {\bf B 135} (1984) 455.


\bibitem{Brodsky} S.J. Brodsky, E. Gardi, G. Grunberg and J. Rathsman, {\em Phys. Rev.} {\bf D 63} (2001) 094017
 [hep-ph/0002065].

\bibitem{Watson} N.J. Watson, {\em Nucl. Phys.} {\bf B 494} (1997) 388 [hep-ph/9606381]; {\em Nucl. Phys.} {\bf B
552} (1999) 461 [hep-ph/9812202]; hep-ph/9912303.

\bibitem{Papavassiliou} J. Papavassiliou, {\em Phys. Rev.} {\bf D 62} (2000) 045006 [hep-ph/9912338]; {\em Phys. Rev. Lett.}
 {\bf 84} (2000) 2782 [hep-ph/9912336].

\bibitem{CMW} S. Catani, G. Marchesini and B. R. Webber,  {\em Nucl. Phys.} {\bf B 349} (1991) 635.

\bibitem{Grunberg-FP} G. Grunberg, {\em Phys. Lett.} {\bf B 372} (1996) 121 [hep-ph/9512203]; in DIS96 [hep-ph/9608375].

\bibitem{D-U} Yu.L. Dokshitzer and N.G. Uraltsev, {\em Phys. Lett.} {\bf B 380} (1996) 141 [hep-ph/9512407].


\bibitem{Vermaseren}  J. A. M. Vermaseren, S.A. Larin and T. van Ritbergen {\em Phys. Lett.} {\bf B 405} (1997) 327
[hep-ph/9703284].

\bibitem{Rafael} S. Peris and E. de Rafael, {\em Nucl. Phys.} {\bf B 500} 
(1997) 325 [hep-ph/9701418].

\bibitem{Gru-condensate} G. Grunberg, {\em Phys. Lett.} {\bf B 325} (1994) 441.

\bibitem{G-qed} G. Grunberg, {\em Phys. Lett.} {\bf B 349} (1995) 469.






 






 



\end{thebibliography}
\end{document}